# Quantum Approximate Optimization Algorithm Parameter Prediction Using a Convolutional Neural Network

**Ningyi Xie[1], Xinwei Lee[1], Dongsheng Cai[1], Yoshiyuki Saito[2], and Nobuyoshi Asai[2]**

[1] University of Tsukuba, Ibaraki Prefecture, Japan
[2] University of Aizu, Fukushima Prefecture, Japan

Ningyi Xie: nyxie@cavelab.cs.tsukuba.ac.jp

**Abstract**. The Quantum approximate optimization algorithm (QAOA) is a quantum-classical hybrid algorithm aiming to produce approximate solutions for combinatorial optimization problems. In the QAOA, the quantum part prepares a quantum parameterized state that encodes the solution, where the parameters are optimized by a classical optimizer. However, it is difficult to find optimal parameters when the quantum circuit becomes deeper. Hence, there is numerous active research on the performance and the optimization cost of QAOA. In this work, we build a convolutional neural network to predict parameters of depth $p+1$ QAOA instance by the parameters from the depth $p$ QAOA counterpart. We propose two strategies based on this model. First, we recurrently apply the model to generate a set of initial values for a certain depth QAOA. It successfully initiates depth 10 QAOA instances, whereas each model is only trained with the parameters from depths less than 6. Second, the model is applied repetitively until the maximum expected value is reached. An average approximation ratio of 0.9759 for Max-Cut over 264 Erdős–Rényi graphs is obtained, while the optimizer is only adopted for generating the first input of the model.

## 1. Introduction

The Quantum approximate optimization algorithm (QAOA) [9] is one of the most notable algorithms in the near-term Noisy Intermediate Scale Quantum (NISQ) era [19]. The advantages of solving combinatorial optimization problems have been discussed in [7][16].

The general process of the QAOA is optimizing a pre-defined objective function by tweaking its parameters contained in the quantum circuit with the help of a classical optimizer. A deeper QAOA circuit is considered promising to derive a more accurate approximation. However, two additional parameters are introduced as the depth increases by one, which increases the difficulty of optimization. Specifically, the optimizer requires more iterations and may be trapped by local optimums easily.

Recently, many strategies are proposed to improve the performance of QAOA with fewer iterations. Among them, those strategies using machine learning show promising power in the cost of searching for appropriate parameters [2][17][3]. However, current machine learning-based methods have defects in universality. Numerous models are required for different problem sizes or QAOA depths.

In this work, we use convolutional neural networks which are largely adopted in image-processing tasks [8][11][15], since the pixels in the images share similar properties with the QAOA parameters. First, pixels are continuous values within an interval, which is the same as QAOA parameters.

Meanwhile, previous studies suggest that there are some correlations between the optimal QAOA parameters [7][21][2][14], just as the adjacent pixels are also correlated. We name the proposed model Parameter-to-Parameter Convolutional Neural Network (PPN)[1], as the model learns mappings between parameters from QAOA with different depths. The proposed PPN can take the output as the input repeatedly to yield prediction for arbitrary depth QAOA. Based on this, we build two strategies.

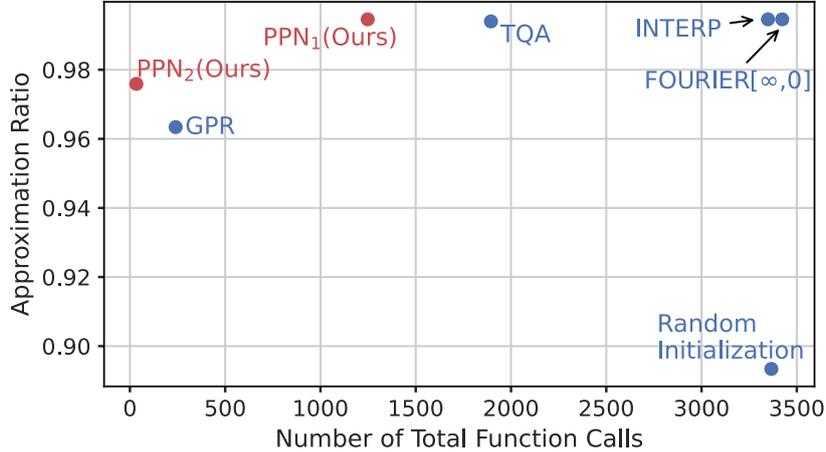

**Figure 1.** The performance and the cost, a visualization of Table 1. Our proposed strategies are highly efficient compared with previous strategies.

To benchmark our strategies, we adopt a typical combinatorial optimization problem Max-Cut and define a problem set. The problem set follows [2], containing 330 8-node Erdős–Rényi graphs with 0.5 edge probability, where 264 graphs are adopted for the test and 66 graphs are used to create the training dataset. All Erdős–Rényi graphs are generated by the function in the NetworkX package [10].

We compare our proposed strategies with the previous work as shown in Figure 1, both strategies improve the efficiency of the QAOA.

The contributions of this study are as follows:
- We significantly reduce the call number of quantum circuits compared to the depth-by-depth strategies [21] and also to state-of-the-art strategy TQA initialization [20].
- We propose a more universal QAOA parameter predictor that uses a convolutional neural network.

The paper is organized as follows. Section 2 provides the background of the QAOA algorithm and Max-Cut problem. In Section 3, we describe previous research and strategies. We introduce the proposed models and strategies in Sections 4 and 5. Section 6 compares our strategies with others, numerically. Finally, we conclude this paper and discuss the future work in Section 7.

## 2. Background

*2.1. The Quantum approximate optimization algorithm*

For the combinatorial optimization problem, the goal is to find a solution $z$ that makes the objective function $C(z)$ maximum or minimum, where $z$ belongs to a feasible region $D$.

In the QAOA [9], the solutions are obtained by measuring the parameterized quantum state. To yield appropriate solutions, the general approach is to optimize the fixed-depth QAOA objective function which is the expected value of $C(z)$:

---

[1] The implementation is available at https://github.com/NingyiXie/Parameter-to-Parameter-Convolutional-Neural-Network.

$$F_{QAOA}\left(\vec{\gamma}^{(p)},\vec{\beta}^{(p)}\right) = \left\langle \Psi\left(\vec{\gamma}^{(p)},\vec{\beta}^{(p)}\right) \middle| H_c \middle| \Psi\left(\vec{\gamma}^{(p)},\vec{\beta}^{(p)}\right) \right\rangle$$
$$= \sum_{z=0}^{2^n-1} P\left(|z\rangle \middle| \vec{\gamma}^{(p)},\vec{\beta}^{(p)}\right) C(z), \quad (1)$$

where $n$ denotes the problem size that $D \subseteq \{z \in \mathbb{Z}, 2^n - 1 \geq z \geq 0\}$, $|z\rangle$ denotes the basis state that encodes a solution, $P\left(|z\rangle \middle| \vec{\gamma}^{(p)},\vec{\beta}^{(p)}\right)$ is the probability of measuring $z$, and the cost Hamiltonian is denoted by $H_C$ that satisfies $\langle z|H_C|z\rangle = C(z)$. The parameter set of depth $p$ QAOA consists of $\vec{\gamma}^{(p)}$ and $\vec{\beta}^{(p)}$, the prepared state $\left|\Psi\left(\vec{\gamma}^{(p)},\vec{\beta}^{(p)}\right)\right\rangle$ can be formulated as:

$$\left|\Psi\left(\vec{\gamma}^{(p)},\vec{\beta}^{(p)}\right)\right\rangle = \prod_{j=1}^{p} e^{-i\beta_j^{(p)} H_B} e^{-i\gamma_j^{(p)} H_C} |+\rangle^{\otimes n}, \quad (2)$$

where $H_B$ is a mixer Hamiltonian that $H_B = \sum_{j=1}^{n} \sigma_j^x$ generally. Here, $\sigma_j^x$ denotes the Pauli-X operator acting on the $j^{th}$ qubit.

The performance of QAOA can be benchmarked with the approximation ratio $\alpha$:

$$\alpha = \frac{F_{QAOA}\left(\vec{\gamma}^{(p)},\vec{\beta}^{(p)}\right)}{C(z^*)}, \quad (3)$$

where $z^*$ denotes the optimal solution.

*2.2. The Max-Cut problem*

The Max-Cut problem is one of the typical combinatorial optimization problems. The goal of the problem is to cut the nodes of a graph into two sets, making the edges between nodes from the two sets as many as possible. It is mathematically equivalent to finding the maximum of the following function:

$$C(z) = \frac{1}{2} \sum_{(j,k) \in E} w_{jk} \left(1 - (-1)^{z_j}(-1)^{z_k}\right), \quad (4)$$

where $z$ represents a bit string $z_1 z_2 \ldots z_n$, $E$ is the edge set of the given graph, and $w_{jk}$ is the weight of $(j,k)$.

The cost Hamiltonian $H_C$ of the Max-Cut problem is constructed with Pauli-Z operators as follows:

$$H_C = \frac{1}{2} \sum_{(j,k \in E)} w_{jk} \left(I - \sigma_j^z \sigma_k^z\right), \quad (5)$$

where $\sigma_j^z$, $\sigma_k^z$ acts on $j^{th}$ and $k^{th}$ qubits, respectively.

In this paper, we simply set the weight of all edges to be 1. Hence, we can bound each $\gamma$ and $\beta$ in intervals $[0, \pi)$ and $\left[0, \frac{\pi}{2}\right)$ [14].

## 3. Related Work

Previous studies [7][21][2][14] reveal that the optimal QAOA parameters have specific patterns. When increasing parameter indexes, the optimal $\gamma$ increases, while $\beta$ decreases. In addition, the optimal $\gamma$ and $\beta$ with the same index deviate slightly as the depth $p$ increases.

Parameters with a higher index at a higher depth are associated with those at the lower depths with lower indexes. This fact leads to some depth-by-depth strategies that use the optimal parameters from lower-depth QAOA to generate the initial values by interpolation [21] or extrapolation [14]. Another depth-by-depth method entitled FOURIER [21] encodes $\vec{\gamma}$ and $\vec{\beta}$ using discrete sine/cosine transforms. Since only the low-frequency components of the transformation are important, this strategy allows us to reduce the number of parameters.

The Trotterized quantum annealing (TQA) initialization [20] inspired by quantum annealing [12][6] is used to initiate the parameters of target depth QAOA directly. According to this method, relationships between parameters are determined as follows: $\gamma_i = (i/p_t)\delta t$, $\beta_i = (1 - i/p_t)\delta t$, where $i$ is the parameter index, $p_t$ denotes the target depth, and $\delta t$ is a pre-defined variable depending on the type of the problem graph. This approach avoids depth-by-depth optimization and thus, saves the computational cost, relatively.

At the same time, some machine learning-based strategies are also proposed. Alam et al. [2] adopt regression models to predict the higher-depth parameters from the lower ones. Moussa et al. [17] apply a clustering algorithm to the encoding of the training problems. After clustering, the best parameter set is selected from the cluster centers and used for the given problem. Amosy et al. [3] adopt a neural network to generate the parameter set from the encoding of the given problem directly.

However, compared with the non-machine learning methods, those machine learning-based methods are trained in the specific target depth [2][17], or work for a specific problem size [3]. Hence, we currently propose a more universal model that allows predicting parameters for any depth or problem size.

## 4. The Model

In this section, we describe the architecture of the proposed Parameter-to-Parameter Convolutional Neural Network (PPN) model and the loss function.

*4.1. Network architecture*

The goal of the PPN is to learn a mapping function $F_{PPN}$ from the parameters of depth $p$ QAOA $\left(\vec{\gamma}^{(p)}, \vec{\beta}^{(p)}\right)$ to the parameters of depth $p+1$ QAOA $\left(\vec{\gamma}^{(p+1)}, \vec{\beta}^{(p+1)}\right)$. We set $\left(\vec{\gamma}^{(p)}, \vec{\beta}^{(p)}\right)$ and $\left(\vec{\gamma}^{(p+1)}, \vec{\beta}^{(p+1)}\right)$ in tensors of size $1 \times 2 \times p$ and $1 \times 2 \times (p+1)$, respectively. We normalize the input parameters to the interval $[0,1)$, and the output of PPN needs to be recovered to the original intervals that $\gamma \in [0, \pi)$ and $\beta \in \left[0, \frac{\pi}{2}\right)$.

As shown in Figure 2, the PPN can be decomposed into three parts: the up-sampling, the residual blocks, and the down-sampling, which can be represented by the three functions $f_1$, $f_2$ and $f_3$ as follows: $F_{PPN} = f_3\left(f_2\left(f_1\left(\vec{\gamma}^{(p)}, \vec{\beta}^{(p)}\right)\right)\right)$. All the parts consist of the convolutional layers. We denote a convolution layer as $Conv(k, m, g)$, where the variables $k$, $m$, $g$ represents the kernel size, the number of filters, and the padding, respectively. The strides for all convolutional layers are assigned as 1. Moreover, Rectified Linear Unit (ReLU) [1] activation functions are applied to the network for the nonlinear mapping. These three parts are explained as follows:

*Up-sampling:* In the first part, the PPN extracts features from the inputted QAOA parameters through two convolutional layers. Each layer is set with $2 \times 2$ kernels and given 1 zero-padding to all four sides of inputs. Meanwhile, the first layer expands the input into 16-dimension, and the latter expands the dimensionality from 16 to 64. Hence, $f_1\left(\vec{\gamma}^{(p)}, \vec{\beta}^{(p)}\right)$ is obtained with $64 \times 4 \times (p+2)$ size at last.

*Residual blocks:* In the computer vision tasks, networks with residual structure exhibit superior performance [11][15]. Inspired by these previous works, we build $D$ residual blocks with convolutional layers and ReLU as the "body" of the PPN. Since all convolutional layers are constructed with $3\times 3$ kernel size and 1 zero-padding, the feature maps $f_2\left(f_1\left(\vec{\gamma}^{(p)},\vec{\beta}^{(p)}\right)\right)$ maintain the tensor of $64\times 4\times (p+2)$ size in this stage.

*Down-sampling:* In this part, we simply use one convolution layer with one $64\times 3\times 2$ filter to down-sample and shrink the feature maps into $1\times 2\times (p+1)$ size $f_3\left(f_2\left(f_1\left(\vec{\gamma}^{(p)},\vec{\beta}^{(p)}\right)\right)\right)$, which is the output of the PPN.

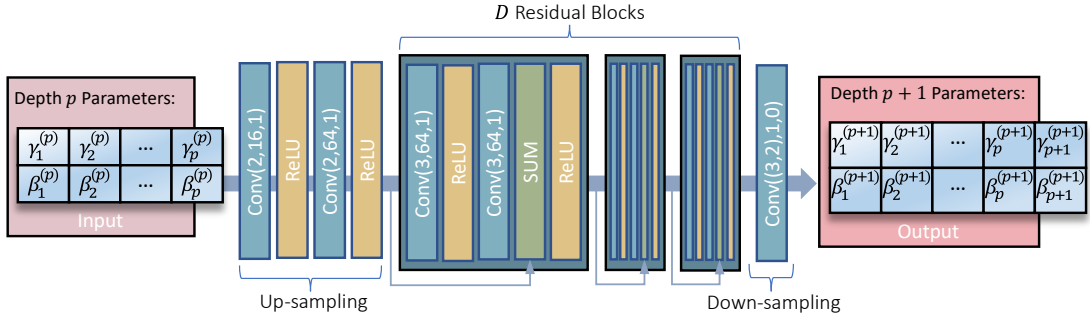

**Figure 2.** Architecture of the Parameter-to-Parameter Convolutional Neural Network (PPN).

*4.2. Loss function*

The output of the PPN is used as the input to predict the parameters of the next depth. Therefore, the output of each stage needs to be close to the ground truth, which are optimal parameters of corresponding depth QAOA.

If the PPN predicts parameters sets from the depth $s$, the output of the $t^{th}$ can be defined as:

$$F_{PPN}^t\left(\vec{\gamma}^{(s)},\vec{\beta}^{(s)}\right)=\left(F_{PPN}\circ F_{PPN}\circ\ldots\circ\right)F_{PPN}\left(\vec{\gamma}^{(s)},\vec{\beta}^{(s)}\right), \quad (6)$$

where $\circ$ denotes a function composition and $F_{PPN}^t$ is $t$ times the function composition of $F_{PPN}$.

The training dataset contains optimal parameters from $s$ to $s+T$ depth QAOA for $N$ problems can be written as $\left\{\left(\left(\vec{\gamma}^{*(s)},\vec{\beta}^{*(s)}\right)^{(i)},\left\{\left(\vec{\gamma}^{*(s)},\vec{\beta}^{*(s)}\right)^{(i)},p=s+(1,2,\ldots,T)\right\}\right),i=1,2,\ldots,N\right\}$. Here, we mark the optimal value with $*$. There are $T$ objectives to minimize. Hence, we have the loss function, which is the averaged mean squared error as follows:

$$\mathcal{L}_{tr}(\Theta)=\frac{1}{NT}\sum_{i=1}^N\sum_{t=1}^T\left\|F_{PPN}^t\left(\left(\vec{\gamma}^{*(s)},\vec{\beta}^{*(s)}\right)^{(i)}\right)-\left(\vec{\gamma}^{*(s+t)},\vec{\beta}^{*(s+t)}\right)^{(i)}\right\|^2, \quad (7)$$

where $\Theta$ denotes the parameter set of the PPN. The loss function is minimized to find the optimal parameter set $\Theta^*$.

## 5. The Strategies

As our proposed PPN can take the parameters of arbitrary depth QAOA as the input and output the parameters of any desired depth QAOA as the prediction, we can devise the strategies flexibly. In this paper, for example, we propose two strategies for QAOA based on the proposed PPN. They both only require the optimal parameters from depth 1 QAOA. For the depth 1 QAOA, we refer to the strategy proposed in [17], making a parameter recommended list, based on a linear-regression result. In this section, we describe these three strategies.

*5.1. Initial strategy*

To solve a combinational problem, we first optimize the depth 1 QAOA circuit to derive the optimal parameter set $\left(\vec{\gamma}^{*(1)}, \vec{\beta}^{*(1)}\right)$ and input it to a trained PPN. Then, we take the output of the PPN as the input, repeatedly, until output $\left(\vec{\gamma}^{*(p_t)}, \vec{\beta}^{(p_t)}\right)$, which is the prediction for target depth $p_t$. Finally, we run QAOA again, with the initial value $\left(\vec{\gamma}^{*(p_t)}, \vec{\beta}^{(p_t)}\right)$. We name this process as $PPN_1$.

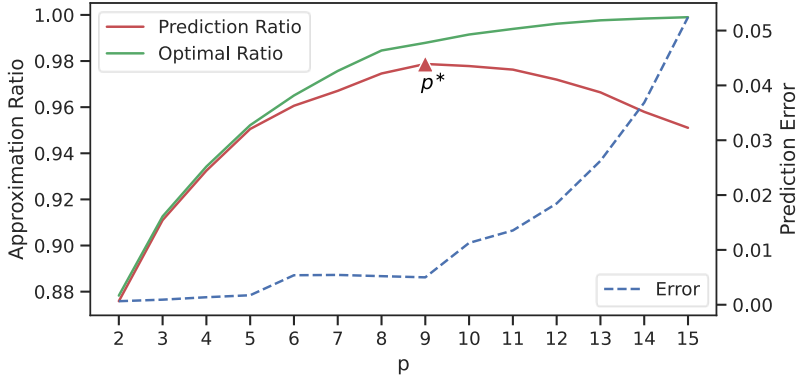

**Figure 3.** The optimal and predicted approximation ratios, and the prediction error. The given problem is the Max-Cut of an 8-node Erdős–Rényi graph with 0.5 edge probability.

*5.2. Depth search strategy*

Optimal approximation ratio is a non-decreasing function of QAOA depth. In fact, the optimal approximation ratio is often a monotonically increasing function of depth for many Max-Cut instances. Hence, even if the parameter set of a higher-depth QAOA instance is not optimal, the expected value $F_{QAOA}$ can be larger than the optimum of a lower-depth instance. In this strategy, we utilize PPN to yield the parameters for QAOA as deep as possible.

To reduce the optimizing iterations, we adopt the optimal parameters of depth 1. Then, we find the optimal output generated by the PPN, recurrently. It can be simplified as the following problem:

$$\max_p F_{QAOA}\left(F_{PPN}^{p-1}\left(\vec{\gamma}^{*(1)}, \vec{\beta}^{*(1)}\right)\right). \quad (8)$$

We find the predicted approximation ratio grows with $p$ at first, then decrease after it hits the maximum, as shown in Figure 3, since the prediction error has a rapid rising after the maximum point of the predicted approximation ratio. This may be caused by the limitation of the training data, which only contains parameters from the lower-depth QAOA. The prediction error is defined by the distance square:

$$\mathcal{L}_{pred}\left(\left(\vec{\gamma}^{(p)}, \vec{\beta}^{(p)}\right), \left(\vec{\gamma}^{*(p)}, \vec{\beta}^{*(p)}\right)\right) = \sum_{i=1}^{p}\left[\left(\gamma_i^{(p)} - \gamma_i^{*(p)}\right)^2 + \left(\beta_i^{(p)} - \beta_i^{*(p)}\right)^2\right], \quad (9)$$

where all parameters are normalized into $[0,1)$.

Due to this error behavior, we can find the optimal depth $p^*$ (i.e., the depth corresponding to the maximum point) easily using Algorithm 1, which we marked as $PPN_2$ in this paper. Once parameters are predicted, we test it using $F_{QAOA}$, if the expected value increases, we take the current parameters to predict the next depth by the PPN. On the other hand, if the expected value decreases, we stop and output the previous results. Hence, $PPN_2$ only needs $p^*$ extra function calls, as it calls function once per depth (from depth 2 to depth $p^*+1$).

**Algorithm 1:** Depth search strategy ($PPN_2$)
**Input:** Problem, $F_{PPN}$.

1. $bestOutput \leftarrow 0$;
2. $(\gamma^{*(1)}, \beta^{*(1)}) \leftarrow$ execute depth 1 QAOA;
3. $\Phi \leftarrow (\gamma^{*(1)}, \beta^{*(1)})$;
4. $tempOutput \leftarrow F_{QAOA}(\gamma^{*(1)}, \beta^{*(1)})$;
5. **while** $tempOutput > bestOutput$ **do**
6.      $bestOutput \leftarrow tempOutput$;
7.      $\Phi \leftarrow F_{PPN}(\Phi)$;
8.      $tempOutput \leftarrow F_{QAOA}(\Phi)$;
9. **end**

*5.3. Linear regression for depth 1 QAOA*

We find that optimal parameters of the depth 1 QAOA $\gamma_1^{*(1)}$ and $\beta_1^{*(1)}$ show a strong positive correlation as shown in Figure 4. The Pearson correlation coefficient [5] between $\gamma_1^{*(1)}$ and $\beta_1^{*(1)}$ from the training dataset is about 0.934. Therefore, we apply the linear least-squares regression to the $\gamma_1^{*(1)}$ and $\beta_1^{*(1)}$ from the training dataset. Since the optimal position is close to the regression line, we pick several points on the regression line as a recommended parameter list. For a given problem, the whole recommended list is traversed to find the parameter set corresponding to the maximum $F_{QAOA}$. Then, the chosen parameter set is adopted as the initial value of the depth 1 QAOA.

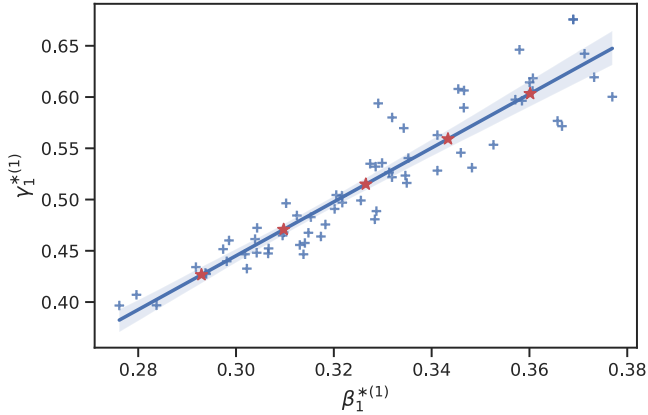

**Figure 4.** The points marked "+" are optimal positions of depth 1 QAOA in the training data. This figure shows the strong linear correlation between $\gamma_1^{*(1)}$ and $\beta_1^{*(1)}$. We use this linear regression to obtain the recommended parameter list (red points).

**6. Experiments**

We adopt the approximation ratio to evaluate the performance of these strategies. And we follow [2][3] to use the number of $F_{QAOA}$ function calls to evaluate the computational cost.

*6.1. Implementation details*

We train a PPN with 4 residual blocks. The training datasets contain optimal parameters from depth 1 to depth 5 QAOA of 66 Max-Cut problems as mentioned in Section 1, which are obtained by a depth-by-depth approach, named the bilinear strategy [14]. During the training, we choose the ADAM [13] as the optimizer. The learning rate is initialed as $1e-5$, and batch size is set as 11 for the first 3000 epochs. Then, we use a batch size of 6 and reduce the learning rate to $1e-6$ for fine-tuning the models with 1000 epochs. We implement this with Pytorch [18] on an NVIDIA 3080Ti GPU.

We test random initialization, INETRP [21], FOURIER[∞, 0] [21], gaussian process regression (GPR)-based strategy [2], TQA initialization [20], and our strategies on the test set, which consists of

264 Max-Cut problems. For INETRP, FOURIER[∞, 0], GPR-based strategy, and our PPN-based strategies, we find the optimal parameters of the depth 1 QAOA using the strategy mentioned in Section 5.3. The $\delta t$ of TQA is pre-defined as 0.625. The GRR model adopted in the GPR-based strategy shares the same training datasets as ours.

The simulation of the quantum circuits is performed using Qiskit's statevector simulator [4] and L-BFGS-B [22] is adopted as the classical optimizer for QAOA.

*6.2. Results and discussion*

Figure 1 shows the cost and the performance of tested strategies. The strategies in the upper left have high performance with relatively lower costs. Our proposed strategies are more cost-effective than others in current experiments.

In detail, as shown in Table 1, our proposed $PPN_1$ achieves the same approximation ratio as that of INETRP [21] and FOURIER[∞, 0] [21], with less than half of the function call. Compare with TQA initialization [20], though TQA only needs one stage optimization, the limitation of the given initial value causes a larger number of function calls. However, the initial values adopted in $PPN_1$ may be closer to the optimal position that help to reduce about 35% of function calls in contrast to TQA. In addition, TQA initialization does not reach the quasi-optimal solution for all problems. The GPR-based strategy [2] has no capacity to initialize the QAOA circuits whose depth is larger than 5, as the restriction of the training dataset. Thus, the GPR-based strategy requires more training data to achieve the same level of performance as $PPN_1$.

As the QAOA depth is unlimited in $PPN_2$, it has a higher upper bound of the expected value than the GPR-based strategy. As a result, $PPN_2$ achieves a 0.9759 average approximation ratio, which significantly surpasses the GPR-based method with much fewer function calls. This approximation ratio nearly matches the optimal results of depth 6 QAOA in our experiments. However, to optimize a depth 6 QAOA circuit, at least about 300 function calls are required for previous strategies, which is 10 times those of $PPN_2$.

In current experiments, there is no numerical comparison between ours and the clustering-based strategy [17], as the clustering-based strategy requires much less empirical data of optimal parameters. However, the clustering-based strategy also needs optimal parameter sets of a higher depth in the preparing stage to achieve competitive results with ours. A fairer condition is required for the comparison of them.

**Table 1.** The average results of approximation ratio and number of function calls

| Method | Approx. Ratio | Number of Function Calls | | | | | | | | | Extra | Total |
|---|---|---|---|---|---|---|---|---|---|---|---|---|
| | | $p=1$ | $p=2$ | $p=3$ | $p=4$ | $p=5$ | $p=6$ | $p=7$ | $p=8$ | $p=9$ | $p=10$ | | |
| Random Initialization ($p=10$) | 0.8934 | | | | | | | | | | 3366.36 | | 3366.36 |
| INTERP [21] ($p=10$) | 0.9946 | 24.66 | 58.94 | 91.42 | 142.36 | 208.38 | 291.52 | 399.38 | 527.45 | 698.11 | 906.50 | | 3348.72 |
| FOURIER[∞,0] [21] ($p=10$) | 0.9946 | 24.66 | 55.85 | 95.27 | 145.98 | 214.71 | 301.12 | 413.52 | 534.73 | 715.45 | 922.09 | | 3423.38 |
| GPR-based [2] ($p=5$) | 0.9634 | 24.66 | | | | 215.17 | | | | | | | 239.83 |
| TQA Initialization [20] ($p=10$) | 0.9940 | | | | | | | | | | 1894.85 | | 1894.85 |
| $PPN_1$ (Ours) ($p=10$) | 0.9946 | 24.66 | | | | | | | | | 1223.17 | | 1247.83 |
| $PPN_2$ (Ours) | 0.9759 | 24.66 | | | | | | | | | | 9.00 | 33.66 |

## 7. Conclusions and Future Work

In this paper, we propose a deep-learning model for parameter prediction. The model generates the QAOA parameters no matter what the depth or the problem size is, as the proposed model learns the mapping from depth $p$ to $p+1$ QAOA parameters. Based on this property, we offer two strategies, both show a high cost-efficiency in the parameter selection task. The first one averagely reduces 35% of the number of expected value function calls of the state-of-the-art non-machine learning method in depth 10 cases. The second one achieves a 0.9759 average approximation ratio with much fewer function calls.

However, the current model necessitates one set of parameters of QAOA circuit as the first input. A considerable method for improving is predicting the parameter sets from the problem directly. As the structure or data of the combinatorial optimization problem can be described by a graph [23], in future work, we consider using a graph neural network to learn problem embedding and predict the parameter sets from the embedding to avoid depth 1 circuit optimization.